\documentclass[doublecol]{epl2}
\usepackage{graphicx}
\usepackage{dcolumn}
\usepackage{bm}
\usepackage{epsfig,amsmath}

\def\vec#1{{\bf #1}}
\def\ffrac#1#2{\frac{#1}{#2}}

\begin{document}

\title{Low friction and rotational dynamics of crystalline flakes in solid lubrication}


\author{A.~S.~de Wijn$^1$, A.~Fasolino$^1$, A.~E.~Filippov$^2$, and M.~Urbakh$^3$}
\institute{$^1$ Radboud University Nijmegen, Institute for Molecules and Materials, Heyendaalseweg 135, 6525AJ Nijmegen, the Netherlands\\$^2$ Donetsk Institute for Physics and Engineering of NASU, 83144, Donetsk, Ukraine\\$^3$ School of Chemistry, Tel Aviv University, 69978 Tel Aviv, Israel}

\abstract{
Solids at incommensurate contact display low-friction, 'superlubric', sliding.
For graphene flakes on a graphite surface, superlubric sliding is only temporary due to rotation of the flakes from incommensurate to commensurate contact with the substrate.
We examine this rotational channel of friction in
a prototype geometry of meso- and macroscopic solid lubrication.
By molecular dynamics simulations and theoretical arguments we find that two surfaces lubricated by mobile, rotating graphene flakes exhibit stable superlubric sliding
as for ideally incommensurate contacts also when they are covered by randomly oriented pinned graphene patches.
For commensurate surfaces, we find a low friction state at low temperature where incommensurate states are not destroyed by thermal fluctuations.
}

\maketitle

Superlubricity between incommensurate surfaces provides a desired low-friction state essential for the function of small-scale machines and function of solid lubricants.
Vanishing static friction has been first predicted by Aubry and Peyrard~\cite{vanishingstaticfriction,vanishingfriction2} for infinite lattices on an incommensurate periodic potential.
Later, Shinjo and Hirano~\cite{shinjo} predicted that for infinite incommensurate contacts also the {\it kinetic} friction would vanish and called this effect {\it superlubricity}.
Extremely low friction has been observed for small contacts in AFM experiments by Dienwiebel et al.~\cite{Dienwiebel2004} for the sliding of small graphite flakes on {a single surface of} graphite.
However, the superlubric sliding is experimentally found to be only temporary, going over to stick-slip behavior after several scans along the surface~\cite{torqueandtwist}.
Direct molecular dynamics simulations \cite{torqueandtwist} as well as a theoretical analysis of the nonlinear dynamics \cite{onsgraphiteflakes} demonstrated that this increase of friction is due to the rotation of the flakes to a commensurate contact {with the single surface}.
The incommensurate orientations of the flakes are stable but not robust against thermal fluctuations particularly if the center of mass of a flake travels on top of the substrate atoms~\cite{onsgraphiteflakes}.
The rotation of nano-scale crystals such as graphene flakes is the simplest nontrivial issue in lubrication with lamellar solids such as graphite and MoS$_2$.

\begin{figure}
\includegraphics[width=8.5cm]{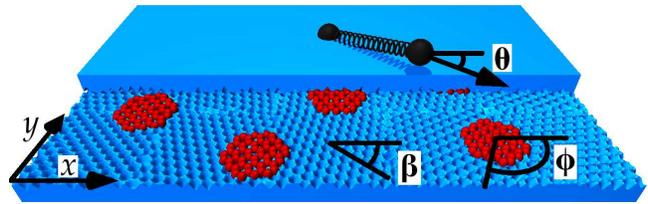}
\caption{
\label{fig:system}
(Color online) The system describes two {infinite} plates with graphene flakes embedded in between.
 The plate surfaces are covered by single-domain or randomly oriented multi-domain graphite layers, as shown for the bottom substrate.
 The top plate is pulled by a spring, the tension in which gives a measure of friction.  The pulling angle $\theta$, relative orientation of lattices on the two plates $\beta$, and flake orientation $\phi$ are indicated.
\looseness-1
}
\end{figure}

In this Letter, we propose a model system that can be used to investigate solid lubrication at both nano and macro scales, and which exhibits stable super-low friction.
The system, shown in Fig.~\ref{fig:system}, consists of two {infinite} plates covered either by single-domain or randomly oriented multi-domain (graphite) layers with (graphene) flakes embedded between them that act as a solid lubricant.
As such, the model contains all key elements
of systems lubricated by lamellar solids, namely a double interface, multiple sliding objects, and rotational dynamics,
which are missing in previous models of friction of a single flake on a single surface~\cite{torqueandtwist,onsgraphiteflakes} and in the models of lubrication with atoms and molecules~\cite{robbinskinetic,robbinsstatic,mueserprl}.
Conversely, we simplify the interaction compared to more precise quantum chemistry calculations such as those of Ref.~\cite{MoS2_2}.
The flakes could be inserted intentionally or result from wear of lamellar solids, as was found in \cite{rapoport2003,martinprb,jolytriblett}.
We show that macroscopic low-friction sliding, expected for single-domain incommensurate contacts, also occurs for disordered plates covered with randomly oriented domains once lubricating flakes are introduced between them.
In fact, although the lubricant flakes readily get into commensurate contact with either of the plates, stick-slip behavior is prevented because the flakes cannot be commensurate with both plates at the same time, contrary to adsorbed flexible chain molecules ('third bodies')~\cite{robbinskinetic,robbinsstatic,mueserprl}.
Besides this, by examining the stability of the superlubric sliding, we find a non trivial mechanism yielding low friction also for commensurate surfaces at low temperature. 

Although our findings are based on a simplified model, considering superlubricity mediated by the flakes gives a perspective to exploit a genuinely atomistic mechanism
at larger length scales.  
Multi-domain graphene layers can be already obtained on macroscopic scales~\cite{Bae} and several other synthesis routes are being currently pursued, making this approach suitable for realistic applications.
Disordered multi-domain surface structures can also be formed spontaneously due to {capillary forces which glue randomly oriented flakes to the sliding surfaces, resulting in a random carpet of flakes rather than a single infinite domain.  This effect prevents rotation of flakes and ensures persistent incommensurability.}
Our model and the results have important implications for understanding the macroscopic properties of graphite and other solid lamellar lubricants, which are undoubtedly the most common solid lubricants~\cite{singer,rapoport}.
Transmission electron microscopy (TEM) observations on MoS$_2$ and WS$_2$~\cite{martinprb,jolytriblett}
have indicated that in macroscopic sliding contacts of lamellar solids rotated flakes are created{.}

We consider $n$ rigid hexagonal graphene flakes formed by $N$ atoms each, which are free to move and rotate in the plane between the two plates.
The typical coherence length of graphene, which is about 1$\mu$m, is much larger than the size of a flake, and therefore elastic deformations can be neglected~{\cite{perssontosatti,Sokoloff2000,mueserstructurallubricity}}.
The coordinates of a flake $i$ are defined by the center-of-mass position $\vec{r}_i$ and the orientation $\phi_i$ with respect to the $x$-axis.
In the numerical simulations, we have taken the total number of embedded atoms $nN=3456$, which corresponds to sufficiently large number of flakes.
Further increase of their number does not affect essentially the friction response.

In order to mimic the commonly used experimental frictional setup, we consider a model where one of the surfaces is stationary, while the other is pulled by a spring coupled to a support moving at constant velocity $\vec{V}$.
We consider friction between atomically smooth surfaces, which is the typical configuration in surface force apparatus (SFA) experiments where the scale of the roughness is larger than the size of the flakes so that it does not influence the translation and rotation motion of the flakes, and the mechanism described in this Letter contributes significantly to the friction.

Each flake atom is
subjected to a potential due to the interaction with each of the two surfaces that  consist either of single domains or of patches with random orientation. The potential function has hexagonal symmetry
\begin{align}
 U(x,y) = - U_0 \left(2 \cos\ffrac{2 \pi x'}{\lambda_1} \cos\ffrac{2\pi y'}{\lambda_2} + \cos\ffrac{4\pi y'}{\lambda_2}\right)~,
\end{align}
where $(x',y')= (x \cos \beta + y \sin \beta, - x \sin\beta + y \cos\beta)$ is the position of a flake atom on a graphite domain with orientation $\beta$.
The parameters $\lambda_1=0.246$~nm, $\lambda_2= \sqrt{3} \lambda_1$, and $U_0=2.26$~meV produce the structure of graphite with a corrugation amplitude of $10.17$~meV.

The dynamics of the system is modelled using Lange\-vin equations
which include random forces $\eta_i(t)$ and $\tilde{\eta}_i(t)$ and random torques $\zeta_i(t)$ and $\tilde{\zeta}_i(t)$, to describe the effects of thermal fluctuations.
Tildes are used to indicate quantities related to the moving plate.
The equation of motion for the position $\vec{R}$ of the moving plate is coupled to those of the flakes as
\begin{align}
 M \ddot{\vec{R}} & = \sum_{i=1}^n \left[-  \tilde{\vec{F}}_{\mathrm{flake}}(\vec{r}_i-\vec{R},\phi_i) + \gamma N m (\dot{\vec{r}}_i-\dot{\vec{R}})\right]\nonumber\\
& \phantom{=}\phantom{\sum[} - \kappa (\vec{R} - \vec{V} t) - \sum_{i=1}^n \tilde{\eta}_i(t)~,\displaybreak[1]\\
 N m \ddot{\vec{r}}_i & = \vec{F}_{\mathrm{flake}}(\vec{r}_i,\phi_i) + \tilde{\vec{F}}_{\mathrm{flake}}(\vec{r}_i-\vec{R},\phi_i)\nonumber\\
& \phantom{=}\null -\gamma N m \dot{\vec{r}}_i -\gamma N m (\dot{\vec{r}}_i-\dot{\vec{R}})+\eta_i(t)+\tilde{\eta}_i(t) ~,\displaybreak[1]\\
 I \ddot{\phi}_i & = T_{\mathrm{flake}}(\vec{r}_i,\phi_i) + \tilde{T}_{\mathrm{flake}}(\vec{r}_i-\vec{R}, \phi_i)\nonumber\\
& \phantom{=}\null - 2 \gamma I \dot{\phi}_i + \zeta_i(t)+ \tilde{\zeta}_i(t)~,
\end{align}
where $M$ is the mass of the driven plate and $m$ is the mass of a carbon atom, $I$ is the moment of inertia of a flake, $\vec{F}_\mathrm{flake}(\vec{r},\phi)$ [$\tilde{\vec{F}}_\mathrm{flake}(\vec{r},\phi)$] is the force acting on a flake at position $\vec{r}$ relative to the plate and orientation $\phi$ due to interaction of all atoms with the surface of the stationary (moving) plate, and $T_\mathrm{flake}(\vec{r},\phi)$ [$\tilde{T}_\mathrm{flake}(\vec{r},\phi)$] is the corresponding torque.
Similar calculations can be done also for multilayer flakes.  What is essential is the structure of the layers which are in contact with the substrates.

The lateral force measured in friction force experiments $\vec{F}= \kappa (\vec{R} - \vec{V} t)$ is the force acting on the spring when the support is moving at constant velocity $\vec{V}$ and its time average is equal to the friction force $F_\mathrm{friction}=|\langle \vec{F}\rangle|$. 
Molecular dynamics simulations of the system have been performed using a fourth-order Runge-Kutta integration algorithm.
 The precision of the algorithm has been tested by energy-conservation at $T=0$ without driving.
 We point out that if energy conservation in this case is not obeyed sufficiently, the stability of the incommensurate orbits is enhanced, artificially reducing the friction.
 With the fourth-order Runge-Kutta algorithm energy-conservation is ensured on the time scale of our simulations with a time step of 0.02~ps.

 The moving plate of mass $M=nN m$, with $m = 1.99\times 10^{-26}$~kg the mass of a carbon atom, is pulled by a spring with spring constant $\kappa = n N\kappa_0 = 4.15$~nN/nm coupled to a support moving at constant velocity $\vec{V} = 12$~m/s at an angle of $\theta=70^\circ$ with respect to the $x$ axis.
 The domains are $10 \lambda_1$ in diameter and the borders between domains are smoothed out by a sinusoidal function.
 We have checked that a reduction in support velocity by up to three orders of magnitude lower does not qualitatively affect the results.
 We have also verified that the pulling direction and domain size affect the results only marginally.

\begin{figure}
\includegraphics[width=8.5cm]{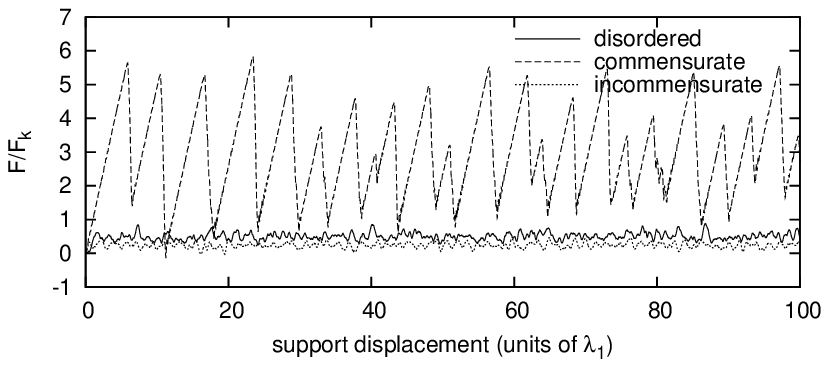}
\caption{
\label{fig:forces}
Lateral force, normalized to $F_k = \kappa\lambda_1$, as a function of the support displacement with $N=24$.  The dashed line is the lateral force for two commensurate surfaces each with a single infinite domain.  The dotted line is the force for two plates with infinite domains at a relative incommensurate orientation of $\beta = 30^\circ$.
 The solid line is for two surfaces covered by random domains with a domain size of 2.46~nm.
 For two commensurate surfaces, the friction shows distinctive stick-slip behavior, whereas for the surfaces with domains and for incommensurate ideal surfaces the forces do not display  stick-slip behavior and are of similar order of magnitude.
}
\end{figure}

In Fig.~\ref{fig:forces}, we show the instantaneous friction forces calculated for the system where the flakes are embedded between fully commensurate plates (dashed line), incommensurate plates (dotted line) and  plates with disordered domains (full line).
One can see that only for fully commensurate plates there is stick-slip behavior resulting in high friction, while for incommensurate and disordered plates, the friction forces are comparable and drastically lower.
This is due to the fact that the flakes cannot lock into commensurate contacts with both plates, not only when the plates are incommensurate, but also in the presence of patches.
Thus the lubrication properties of flakes differ significantly from those of short flexible chains of particles studied in Refs.~\cite{robbinskinetic,robbinsstatic,mueserprl}, where it is shown that, even for incommensurate plates, short chains can lock to both plates and increase the friction.
This feature, together with the fact that wear in lamellar systems leads to creation of lubricant flakes may explain why layered materials~\cite{singer,rapoport} are excellent lubricants.


Under humid conditions, the multi-domain surface structures can be formed spontaneously due to the capillary forces {which fix randomly oriented flakes at the sliding surfaces}, while in vacuum conditions the formation of such structures is less likely. 
{In addition, the water vapor trapped between the surfaces may saturate the bonds on graphite, thus preventing covalent bonding between the graphite surfaces.}

{These} may be the reasons why graphite is such a bad lubricant in vacuum, and needs the humidity of air to perform well.

\begin{figure}
\includegraphics[width=8.5cm]{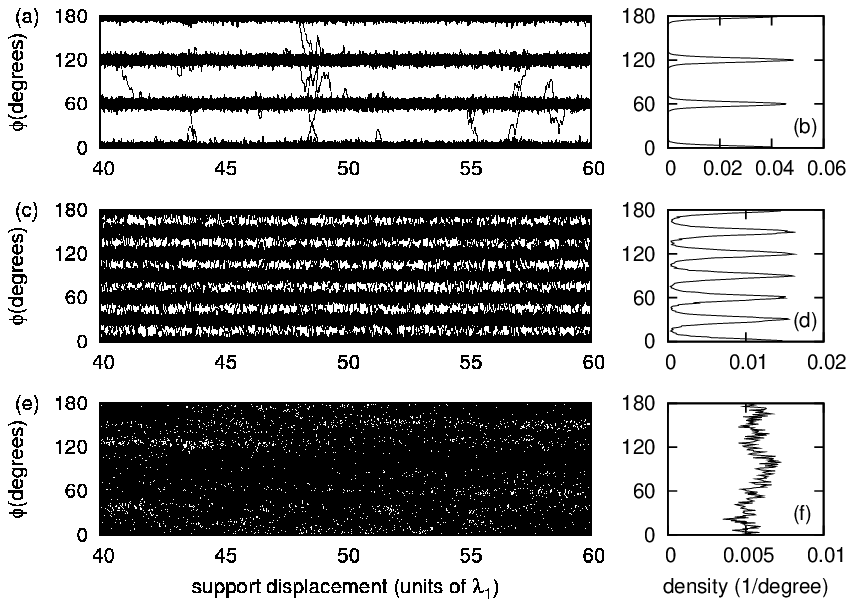}
\caption{
\label{fig:trajectories}
Orientations $\phi_i$ of 24-atom flakes (left) and their distribution functions (right) at $T=77$~K, 
 for two commensurate surfaces [(a) and (b)], two incommensurate surfaces at $\beta = 30^\circ$ [(c) and (d)], and for two multi-domain surfaces [(e) and (f)].
 Note in (a), (b), (c), and (d) that the flakes have orientations commensurate to either of the two graphite surfaces.
 In the case of (c) and (d), the friction is low, because the flakes are not commensurate to both surfaces at the same time.
 This picture is preserved for multi-domain surfaces.
}
\end{figure}

\begin{figure}
\includegraphics[width=8.5cm]{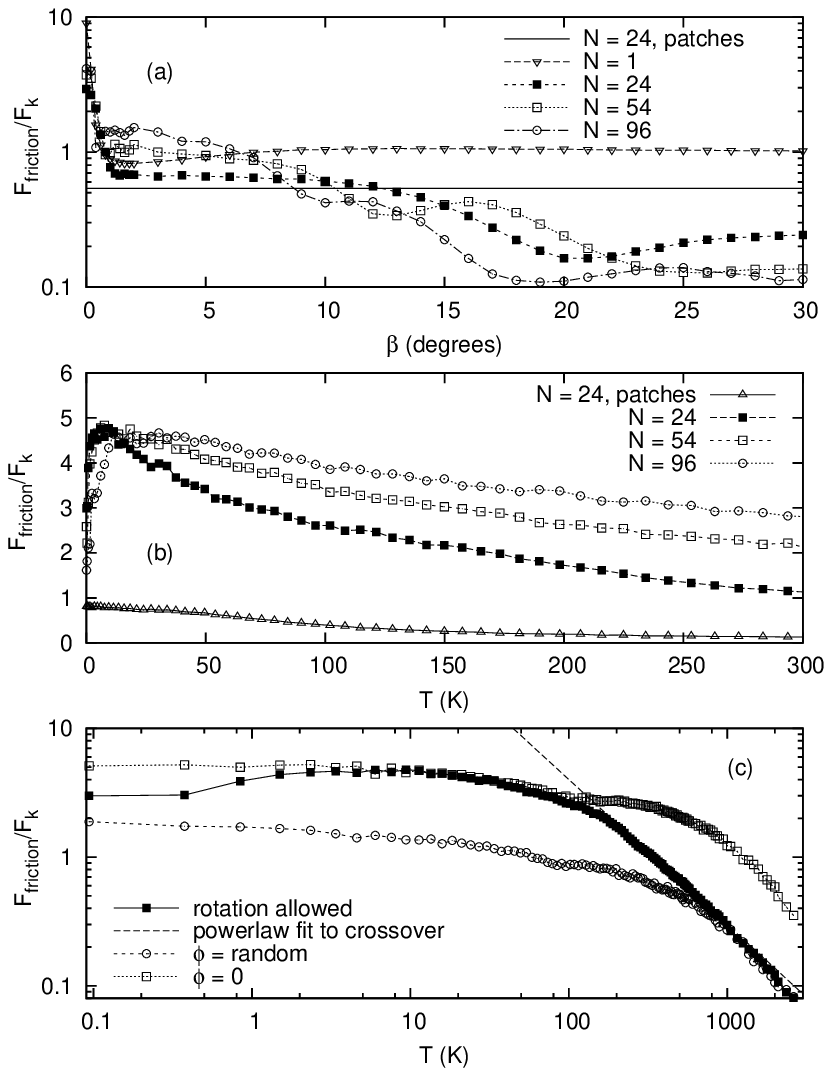}
\caption{
Friction force for various flake sizes as a function of the relative angle of the plates at $T=77$~K (a), as a function of temperature for commensurate plates (b), and as a function of temperature for different rotational dynamics of 24-atom flakes embedded between commensurate plates (c).
 In panel (a), results for 24-atom flakes between two plates with patches are included for comparison and indicated with a solid line at $F_\mathrm{friction}/F_k=0.54$.
 This value is
 close to the average over all angles for single-domain incommensurate plates,
$F_\mathrm{friction}/F_k=0.46\pm0.05$.
\label{fig:FbFT}
}
\end{figure}

Since the main mechanism of friction is determined by the rotation of the flakes from incommensurate (superlubric) to commensurate (stick-slip) orientation, the ensemble of angles $\phi_i$ of the flakes provides information on the dynamics.
We find the surprising result that, in all three cases, for most of the time, the flakes are in commensurate contact with either of the plates.
However, except in the case of commensurate plates, the flakes cannot be commensurate to {\it both} plates at the same time.
It is precisely this frustration that leads to the low value of the friction. 
As shown in Fig.~\ref{fig:trajectories}, bands at commensurate angles are clearly visible for both single-domain commensurate [(a) and (b)] and incommensurate [(c) and (d)] contacts, whereas 
the disordered patches [(e) and (f)] ensure that all orientations are present.
That the latter situation is favorable for low friction can be understood from Fig.~\ref{fig:FbFT}(a), where we show the dependence of friction on the relative orientation of the sliding plate to the stationary one for different flake sizes.
Only in a narrow range of angles $\beta$ close to perfect commensurability the friction is high.
For all other angles, the friction is low, except in the case of single particle lubricants ($N=1$), which cannot create incommensurate contacts~\cite{robbinskinetic,mueserprl}.
For multi-domain surfaces, the relative orientation of the two lattices is random, and therefore, the friction will be equal to an average over all relative angles, resulting in a large reduction with respect to the commensurate contact.

In Fig.~\ref{fig:FbFT}(b) we show that above a given temperature, the friction decreases, as expected.
This decrease sets in later for larger flakes.
Again, patches on both plates lead to much lower friction at all temperatures.

In Fig.~\ref{fig:FbFT}(c), we shown that for commensurate plates, the temperature dependence of the friction exhibits a crossover around T=200~K.  Above this temperature, the friction drops, following a power-law proportional to $T^{-1.13\pm0.04}$.
This crossover can be understood by comparison with the friction calculated by freezing the orientation of the flakes either in the commensurate orientation $\phi_i=0$, or with a random distribution of angles.
At low temperatures,
most flakes have $\phi_i=0$, whereas at higher temperatures, incommensurate orientations come into play and the friction is similar to that of a frozen system with flakes with random orientations.

\begin{figure}[t]
\includegraphics[width=8.5cm]{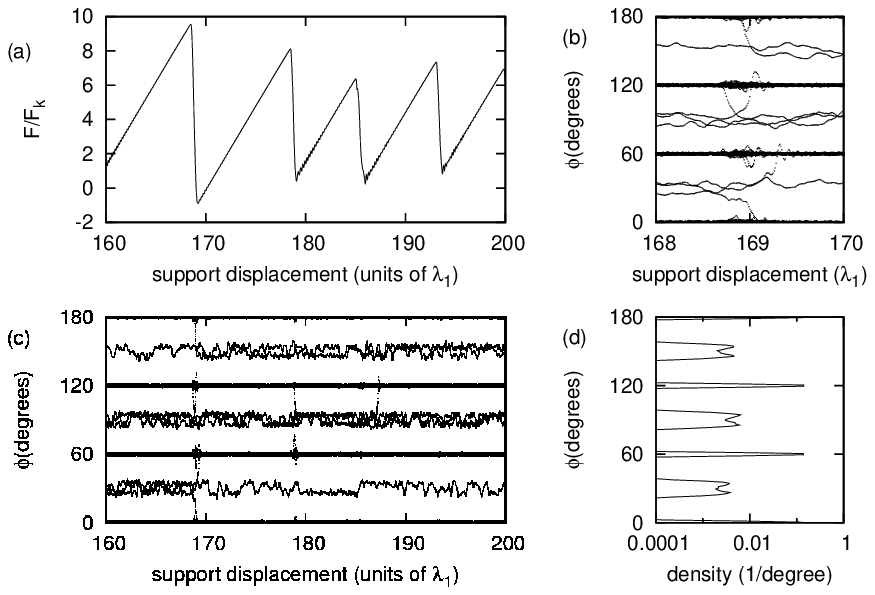}
\caption{
Role of incommensurate orbits at low temperatures at 2.6~K and $\gamma=0.4$/ps.
Lateral force (a) and flake orientation [(b) and (c)] as a function of support displacement.
 The distribution of orientations after long time is shown in (d).
 Rotations from incommensurate to commensurate orientations and vice versa are caused by the energy released during a slip.
 In between slips, the incommensurate orientations survive, because they are in principle stable and 
  they contribute significantly to the total distribution of orientations, and therefore lead to lower friction.
\label{fig:orbits}
}
\end{figure}

Lastly, we discuss the unusual reduction of friction found for all flake sizes at very low temperatures [Fig.~\ref{fig:FbFT}(b) and~\ref{fig:FbFT}(c)].
This effect is due to the existence of stable incommensurate orbits and is shown in more detail in Fig.~\ref{fig:orbits}.
{These orbits can be present in the initial conditions, but they are also activited during slips.}
For two commensurate plates a great deal of energy is released into the system during a slip [see Fig.~\ref{fig:orbits}(a).
This causes some flakes to move into the incommensurate orientations [see Fig.~\ref{fig:orbits}(b) and~(c)], which are not robust against thermal fluctuations.
At high temperatures, the incommensurate orbits decay rapidly to commensurate ones ($\phi=0,60,120^\circ$), and so their effect on friction is negligible ~\cite{torqueandtwist,onsgraphiteflakes}.
At low temperature, however, the decay rate of the incommensurate orbits is so low, that they survive for a long time, significantly reducing the friction.
The distribution of orientations is shown in Fig.~\ref{fig:orbits}(d), where one can see peaks at the commensurate orientation, as in Fig.~\ref{fig:trajectories}(b), and at incommensurate orientations at $\phi=\pm26,34,86,94,146,154^\circ$.
These coincide precisely with the stable incommensurate orientations found for $N=24$~\cite{onsgraphiteflakes}.
The peaks at the incommensurate orientations do not appear in the angle distributions at higher temperature, shown in~Fig.~\ref{fig:trajectories}(b).

For graphite, the corrugation is only about 10~meV, and the low-temperature orbits significantly affect the friction only for temperatures below about 10~K.
However, this effect can become important in a wide range of temperatures for systems with larger potential corrugation, either due to higher normal load or to other substrates with hexagonal symmetry which exhibit larger corrugation, e.g.~BN~\cite{BNcorrugatie1}, which can be grown, for instance, on Rh~\cite{BNmesh}.
In these cases, the frozen-orbit regime could provide an interesting new mechanism for low-friction sliding.
\looseness-1

In summary, we have investigated a simple model which nevertheless contains crucial elements of solid lubrication, in particular the rotational dynamics of the flakes.
We have shown that quasi-superlubric behavior is achieved for the sliding of plates with disordered multi-domain structure lubricated with graphite flakes.
We have shown that rotational dynamics of the lubricant flakes, which destroy superlubric sliding of single flakes on a surface, are not necessarily detrimental in the actual geometry of solid lubrication.

ASdW's work is financially supported by a Veni grant of Netherlands Organisation for Scientific Research (NWO).
 AF's work is part of the research programme of the Foundation for Fundamental Research on Matter (FOM), which is financially supported by NWO.
 MU's work is part of the ESF EUROCORES Program FANAS (ACOF, AFRI, AQUALUBE), and is supported by the Israel Science Foundation (1109/09).


\end{document}